\documentclass[prl,twocolumn,showpacs,superscriptaddress,groupedaddress]{revtex4}  
\usepackage{graphicx}
\usepackage{epstopdf}
\usepackage{amsmath,amsthm,amssymb}
\usepackage{color}
\usepackage{amsfonts}
\usepackage{subfigure}
\usepackage{bm}

\begin{document}

\widetext


\title{Van der Waals explosion of cold Rydberg clusters}


\author{R. Faoro}
\affiliation{Dipartimento di Fisica ``E. Fermi'', Universit\`a di Pisa, Largo Bruno Pontecorvo 3, 56127 Pisa, Italy}
\affiliation{Laboratoire Aim\'e Cotton, CNRS, Universit\'e Paris-Sud, ENS Cachan,  Campus d'Orsay, B\^at 505, 91405 Orsay, France}

\author{C. Simonelli}
\affiliation{Dipartimento di Fisica ``E. Fermi'', Universit\`a di Pisa, Largo Bruno Pontecorvo 3, 56127 Pisa, Italy}
\affiliation{INO-CNR, Via G. Moruzzi 1, 56124 Pisa, Italy}

\author{M. Archimi}
\affiliation{Dipartimento di Fisica ``E. Fermi'', Universit\`a di Pisa, Largo Bruno Pontecorvo 3, 56127 Pisa, Italy}

\author{G. Masella}
\affiliation{Dipartimento di Fisica ``E. Fermi'', Universit\`a di Pisa, Largo Bruno Pontecorvo 3, 56127 Pisa, Italy}

\author{M.M. Valado}
\affiliation{Dipartimento di Fisica ``E. Fermi'', Universit\`a di Pisa, Largo Bruno Pontecorvo 3, 56127 Pisa, Italy}
\affiliation{INO-CNR, Via G. Moruzzi 1, 56124 Pisa, Italy}

\author{E. Arimondo}
\affiliation{Dipartimento di Fisica ``E. Fermi'', Universit\`a di Pisa, Largo Bruno Pontecorvo 3, 56127 Pisa, Italy}
\affiliation{INO-CNR, Via G. Moruzzi 1, 56124 Pisa, Italy}
\affiliation{CNISM UdR Dipartimento di Fisica ``E. Fermi'', Universit\`a di Pisa, Largo Pontecorvo 3, 56127 Pisa, Italy}

\author{R. Mannella}
\affiliation{Dipartimento di Fisica ``E. Fermi'', Universit\`a di Pisa, Largo Bruno Pontecorvo 3, 56127 Pisa, Italy}
\affiliation{CNISM UdR Dipartimento di Fisica ``E. Fermi'', Universit\`a di Pisa, Largo Pontecorvo 3, 56127 Pisa, Italy}

\author{D. Ciampini}
\affiliation{Dipartimento di Fisica ``E. Fermi'', Universit\`a di Pisa, Largo Bruno Pontecorvo 3, 56127 Pisa, Italy}
\affiliation{INO-CNR, Via G. Moruzzi 1, 56124 Pisa, Italy}
\affiliation{CNISM UdR Dipartimento di Fisica ``E. Fermi'', Universit\`a di Pisa, Largo Pontecorvo 3, 56127 Pisa, Italy}

\author{O. Morsch}
\affiliation{Dipartimento di Fisica ``E. Fermi'', Universit\`a di Pisa, Largo Bruno Pontecorvo 3, 56127 Pisa, Italy}
\affiliation{INO-CNR, Via G. Moruzzi 1, 56124 Pisa, Italy}

\date{\today}

\begin{abstract}
We report on the direct measurement in real space of the effect of the van der Waals forces between individual Rydberg atoms on their external degrees of freedom. Clusters of Rydberg atoms with inter-particle distances of around $5\,\mathrm{\mu m}$ are created by first generating a small number of seed excitations in a magneto-optical trap, followed by off-resonant excitation that leads to a chain of facilitated excitation events. After a variable expansion time the Rydberg atoms are field ionized, and from the arrival time distributions the size of the Rydberg cluster after expansion is calculated. Our experimental results agree well with a numerical simulation of the van der Waals explosion.
\end{abstract}

\maketitle


The van der Waals (vdW) forces between polarisable neutral particles plays an important role in a variety of physical phenomena~\cite{Leroy:70,Israelachvili:11}. Those forces are weak, appreciable only at very short interparticle distances and, therefore, difficult to measure directly, as demonstrated by the pioneering experiments on the vdW interactions between an atom and a surface~\cite{Anderson:88,Sandoghdar:92}.  Nevertheless, they play an important role in weak chemical bindings and very often in nanoscale phenomena. The vdW interaction is a key element in the physics of Rydberg atoms where, owing to the large distance of the electron from the nucleus, the polarizabilities are larger than for atoms in the ground state~\cite{Comparat:10, Pfau:12}. For instance, the Rydberg level shifts due to the vdW interactions are responsible for the so-called dipole blockade~\cite{Beguin:13}, in which a Rydberg excitation is strongly suppressed owing to the presence of a previous excitation~\cite{Urban:09,Gaetan:09,Weber:15}. Experimental determinations of the vdW interaction rely on the measurement of observables linked to either internal or external degrees of freedom. The first class is represented by the measurements of the level shifts using optical spectroscopy, as in an early study~\cite{Raimond:81}, or in a very recent experiment using microwave spectroscopy to monitor the early stage repulsion between off-resonantly excited Rydberg atoms~\cite{Teixeira:15}. All the above experiments observed the van der Waals force indirectly through the effect of the interaction on atomic energy levels. The observation of the external degrees of freedom of Rydberg atoms was initiated in~\cite{Amthor:07}, and progress in that direction is presented in~\cite{Thaicharoen:2015}. Here we report the direct observation in real space of the mechanical effect of the vdW force on individual Rydberg atoms by investigating the formation and subsequent explosion of Rydberg clusters. Our results enable further studies of the vdW interaction as well as the reconstruction of the spatial arrangement of excited atoms inside a Rydberg cluster, in analogy with the experiments on the Coulomb explosion of pulse-ionized molecules~\cite{Vager:89,Pitzer:13}.\\
\indent Our experiments, described in~\cite{Malossi:14,Viteau:13}, are performed in magneto-optical traps (MOTs) of $^{87}$Rb atoms with typical sizes around $150\,\mathrm{\mu m}$  and peak densities of up to $10^{11}$ cm$^{-3}$. Two laser beams at 420 nm and 1013 nm are used to excite atoms to the $70\mathrm{S}$ Rydberg state, with the 420 nm laser detuned by 660 MHz from the $6\,\mathrm{P}_{3/2}$ intermediate state (the MOT beams are switched off during this excitation). The laser beam at 1013 nm has a waist of $110\,\mathrm{\mu m}$, while the 420 nm beam is focused to around $6 \,\mathrm{\mu m}$, thus realizing a quasi-1D geometry in which the radial size of the 420 nm laser is comparable to the facilitation radius (see below) and hence the excitation dynamics takes place predominantly along the $x$ propagation direction of the beam (see Fig.~1). The 1013 nm beam lies in the $x-y$ plane at a $45^\circ$ angle with respect to the x-axis, leading to an effective excitation volume (defined by the overlap of the two beams) of approximately $10^{-7}\,\mathrm{ cm^{-3}}$.  The two-photon Rabi frequency for the Rydberg excitation is around 2 MHz. The production based on the seed approach (see the following) of between one and two Rydberg in the atomic cloud  uses a short ($\approx0.5\mathrm{\mu s}$) resonant pulse at small two-photon Rabi frequency ($0.3$ MHz) derived from one of the diffraction orders of the acousto-optic modulator controlling the excitation pulses. After excitation and a variable expansion time the Rydberg atoms are field ionized and accelerated towards a channeltron (see in Fig.~1), where the ion arrival times are recorded and analysed following the approach introduced in~\cite{Krasa:98, Henkel:10}. Our detection efficiency is around $50\,\mathrm{\%}$; for long expansion times the overall signal decreases due to spontaneous decay of the Rydberg state.  \\
 \begin{figure}[htbp]
\begin{center}
\includegraphics[width=7.5 cm]{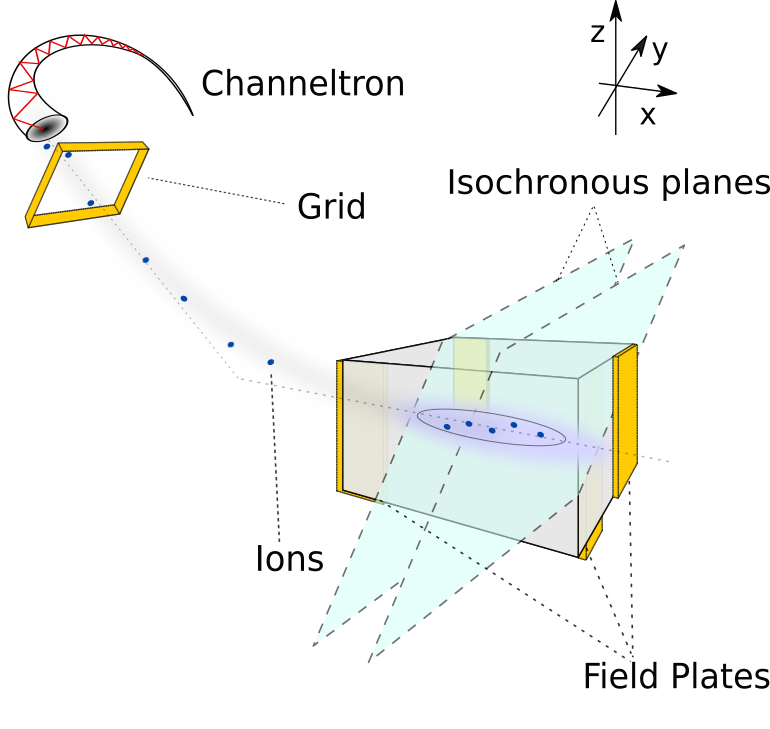}
\caption{Cold atoms in a MOT excited to Rydberg states in an elongated interaction volume (continuous ellipse), and subsequently expand due to the vdW repulsion. After a variable free expansion time they are field ionized by switching on the field plates. Finally the ions (blue dots) are electrically accelerated through a grid towards a channeltron. The isochronous planes defined in the text are schematically represented. }
\label{default}
\end{center}
\end{figure}
\indent In order to be able to determine the size of an expanding cloud of Rydberg atoms over a spatial range of several hundreds of micrometers, we developed a measurement technique based on the arrival times of ions at the channeltron. In our apparatus the distribution of arrival times of ions originating from the interaction volume is determined by the geometry of the ion collection apparatus, in which the ions are accelerated towards the channeltron along a curved path in the $x-z$-plane, as schematically shown in the upper left corner of Fig.~1. Therefore, the ion arrival times depend essentially on their initial positions along the $x$ and $z$ axes. The calibration of our measurement technique was performed by determining the functional dependence of $\Delta t$ on the initial positions of the ions. To do so, we created small MOTs around $150\,\mathrm{\mu m}$ in size at different spatial positions using independent magnetic fields in the three spatial directions.  With the MOT lasers switched on, a short (around $1\,\mathrm{\mu s}$) pulse of only the 420 nm beam created on average a single ion, in order to avoid aberrations in the collection process due to the Coulomb repulsion between the ions. This direct ionization was also chosen in order to avoid Rydberg-Rydberg interaction effects that could have altered the excitation probability and hence the spatial distribution of ions inside the MOT. As verified in \citep{Valado:13}, the  direct ionization  probability is strictly proportional to the local atom density. We measured the average ion arrival times at the channeltron calculated from 400 experimental runs at a given MOT position and thus obtained the dependence of $\Delta t$ on the MOT position as shown in Fig.~2. To a good approximation, over a range of $0.8\,\mathrm{mm}$ the arrival times are are independent of the $y$-position and depend linearly on the $x$- and $z$- positions, with proportionality coefficients $\alpha= -2.31\pm0.08\times10^{-1}\,\mathrm{\mu s}/\mathrm{mm}$ and $\beta=+1.34\pm0.07\times10^{-1}\,\mathrm{\mu s}/\mathrm{mm}$, respectively. Thus our detection process projects the positions of
ions created in an isochronous plane (which contains the $y$-axis and makes an angle of $59.9 \pm1.5^\circ$ with the $x- y$ plane) onto a particular arrival time
at the channeltron, as shown in Fig.~1. We conclude that the ions created in isochronous planes separated by a distance $\Delta s$ give rise to a temporal separation
\begin{equation}
\Delta t=\sqrt{ \alpha^2+\beta ^2} \Delta s
\label{Eq1}
\end{equation} of their arrival times. This dependence was confirmed by a measurement of the $\Delta t$ delays in the case of MOT created varying $z$ and keeping $x$ constant and different from zero (red circles with black borders in Fig.~2). From our data we derive an effective one-dimensional spatial resolution of around $9\,\mathrm{\mu m}$ in the single ion regime, where the Coulomb repulsion between nearby ions during the $\approx 10\,\mathrm{\mu s}$ acceleration time towards the channeltron can be neglected. For the conditions of the experiments reported below, this approximation is valid except for the very early stages ($< 100\,\mathrm{\mu s}$) of the dynamics, and can be used to derive the initial position of the created ions created.\\
\begin{figure}[htbp]
\begin{center}
\includegraphics[width=8cm]{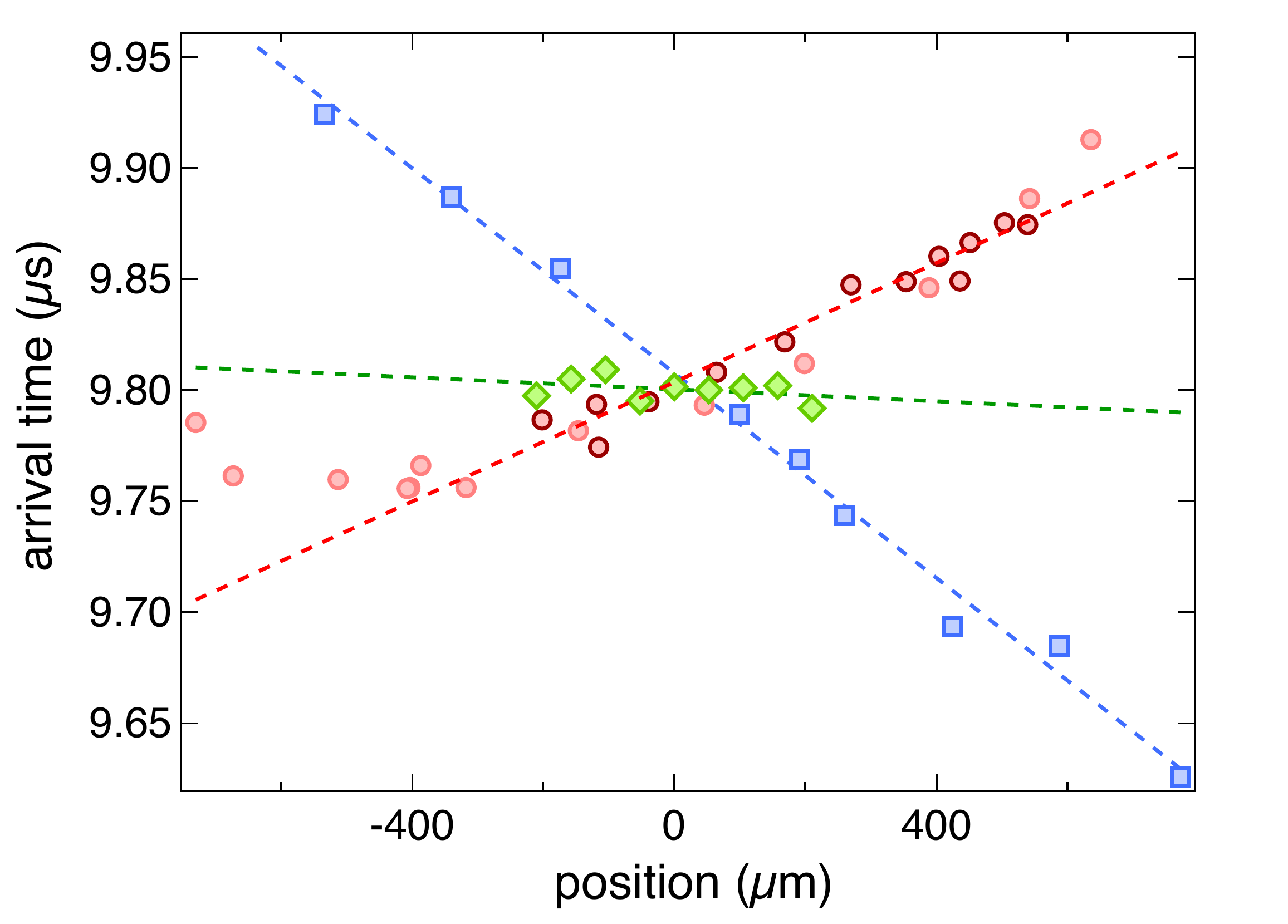}
\caption{Calibration of the arrival time distributions at the channeltron. The center of the arrival time distributions is plotted as a function of the MOT position in the $x$- (blue squares), $y$- (green diamonds) and $z$-directions (red circles); in each case, the MOT was centred at $0$ in the other two directions. The error bars are smaller than the size of the data symbols. The dashed lines are linear fits to the experimental data within the range (-0.4,0.4) $\mathrm{mm}$. The red circles with black borders correspond to a variation in the $z$-position for a fixed $x=460\pm 5 \,\mathrm{\mu m}$ where the arrival times were corrected using the $x$- axis calibration. Their agreement with the $x=0$ data demonstrates that the arrival times are linearly dependent on the $x$ and $z$ positions without cross terms.}
\label{Fig2}
\end{center}
\end{figure}
%

\begin{figure}
\begin{center}
\includegraphics[width=8.7cm]{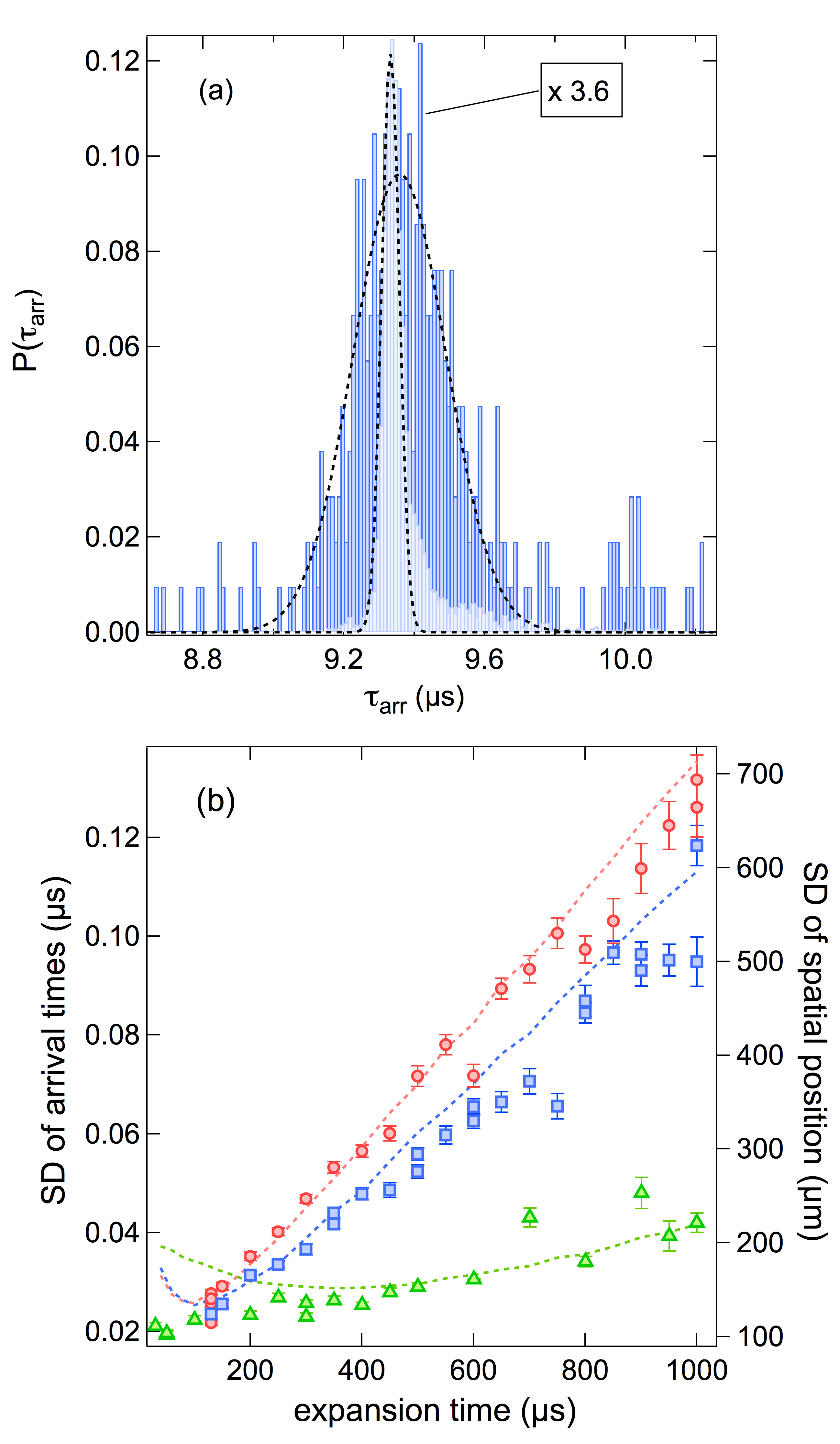}

\end{center}
\caption{(a) Histogram of the probabilities $P$ of  ion arrival times,  $\tau_{arr}$, at $\Delta=80\,\mathrm{MHz}$   for two different expansion times: $130\,\mathrm{\mu s}$ (light blue bars) and $1000\,\mathrm{\mu s}$ (dark blue bars). SD's of $P(\tau_{arr})$ are calculated from Gaussian fits (dashed lines) to the histograms. (b) SD, probing the width of the ion cloud vs the expansion time for different detunings $\Delta$: 0 (green triangles), $55\,\mathrm{MHz}$ (blue squares) and $80\,\mathrm{MHz}$ (red circles).  On the l.h.s axis SD's of the arrival times, and on the r.h.s axis  SD's of the spatial positions, obtained from the previous SD's by using the inversion of Eq.~\eqref{Eq1} are shown. The dashed lines report numerical simulations of the vdW explosion.}
\label{Fig3}
\end{figure}
\indent Two procedures tested in our previous work are applied to the exploration of the ions created following the Rydberg excitation. In order to maximize the van der Waals interactions and, therefore, the interatomic repulsions, we create Rydberg excitations at small interatomic distances by using the facilitated off-resonant excitation explored in~\cite{Amthor:07,Schempp:14,Malossi:14,Teixeira:15} and theoretically treated in~\cite{Lesanovsky:14}. In this scheme, following the initial creation of a Rydberg excitation an additional atom at distance $r_{fac}$ from the first one is also excited if the laser detuning $\Delta$ matches the Rydberg level shift due to the vdW interaction, i.e., $h\Delta=C_6/(r_{fac}^6)$~\cite{Note1}.  Facilitation implies that close to an already excited Rydberg atom all the ground-state atoms located inside a shell of radius $r_{fac}$ and width $\delta r_{fac}= 1/6\times r_{fac}\times \delta/\Delta$ (where $\delta$ is the laser linewidth~\cite{Lesanovsky:14}) can theoretically be excited. The excitation of more than one atom will, however, distort the facilitation volume due to the additional van der Waals interaction between those atoms. For our experimental parameters, the facilitated excitation of more than one atom per facilitation volume is unlikely. For the laser detunings of our investigation the values of $r_{fac}$ and $\delta r_{fac}$ are around$~5\,\mathrm{\mu m}$ and$~6\,\mathrm{nm}$, respectively. The first Rydberg atom created through facilitated excitation, in turn, facilitates the excitation of another atom, and thus a chain reaction is started that continues as long as there are ground state atoms inside the next facilitation shell. Because the facilitation radius is comparable to the transverse dimension of the confining potential the cluster of Rydberg atoms created by the chain process is extended mainly in one dimension, as shown schematically in Fig.~4. Since for the large values of $\Delta$ used in our experiments the spontaneous off-resonant excitation of the initial Rydberg atom needed to start the chain reaction is small, we use a seed technique based on a single resonant pulse leading to the creation of one/two Rydberg excitations within the interaction volume~\cite{Simonelli:14,Simonelli:15}.  After the creation of the seed a 100$\,\mathrm{\mu s}$ pulse produces the facilitated Rydberg excitation. We verify that the number of excited atoms reaches a plateau towards the end of the excitation time by performing a measurement of the mean number of Rydberg atoms by field ionization as in~\cite{Viteau:12} immediately after the end of the excitation pulse. From those measurements we derive that for $\Delta=$ $55\pm1\,\mathrm{MHz}$ and $80\pm1\,\mathrm{MHz}$  the chain reaction of facilitation events produces a quasi-1D Rydberg cluster containing $12$ excited atoms in both cases.\\
\indent We now proceed to use the spatial information contained in the ion arrival times in order to probe the temporal evolution of the size of the Rydberg cloud.  After the above preparation stage the cluster expands freely for up to $1\,\mathrm{ms}$. From our numerical simulations (see below) we find that the vdW energy of pairs of Rydberg atoms initially created at distance $r_{fac}$ is fully converted into kinetic energy after around $5\,\mathrm{\mu s}$, i.e., much less than the the $150 \,\mathrm{\mu s}$  lifetime of the 70S Rydberg state. After the conversion of the vdW energy into kinetic energy excited atoms can, however, decay into lower Rydberg states that are not detected by our electric field ionization process, which only field ionizes and detects Rydberg atoms with principal quantum number $n\gtrsim 50$. By repeating the excitation/detection cycle $400$ times we determine the spatial distribution of the Rydberg cloud.  Here we assume that the spontaneous decay of the Rydberg excitations occurs independently for each atom, so that the arrival time distribution of the non-decayed subset of atoms reflects that of all the initially excited atoms. Typical histograms  of the arrival time distribution are reported in Fig.~3(a).  We measure the standard deviations (SD's) of those distributions as function of the free expansion time, as shown in Fig.~3(b) (left scale). By inverting Eq.\eqref{Eq1}, from the arrival time SD's we derive the spatial width SD's as a function of that time, as shown in Fig.~3(b). For both values of $\Delta$ different from zero  the vdW explosion produces a large spread in the arrival times. In a test experiment at $\Delta=0$ (with $12$ excitations on average), for which no vdW explosion should occur, the cloud expansion at $1\,\mathrm{ms}$ is small and compatible with the expected expansion due the thermal motion. As expected from the strong $r$-dependence, the van der Waals force acts mainly at short interatomic distances and, therefore, the expansion of the cloud is essentially ballistic over the full temporal range. In fact, the large initial acceleration, $\approx 5 \times10^5\,\mathrm{ms^{-2}}$ at $\Delta=80$ MHz and $r_{fac}$, reduces to one percent  at 2$r_{fac}$. At that detuning the theoretically expected overall increase in velocity for two Rydberg atoms with a corresponding initial vdW energy is $\approx 600\,\mathrm{\mu m}/{\mathrm{ms}}$ compared to the average thermal velocity of MOT atoms of $\approx 120\,\mathrm{\mu m}/{\mathrm{ms}}$.\\
\begin{figure}
\begin{center}
\includegraphics[width=6 cm]{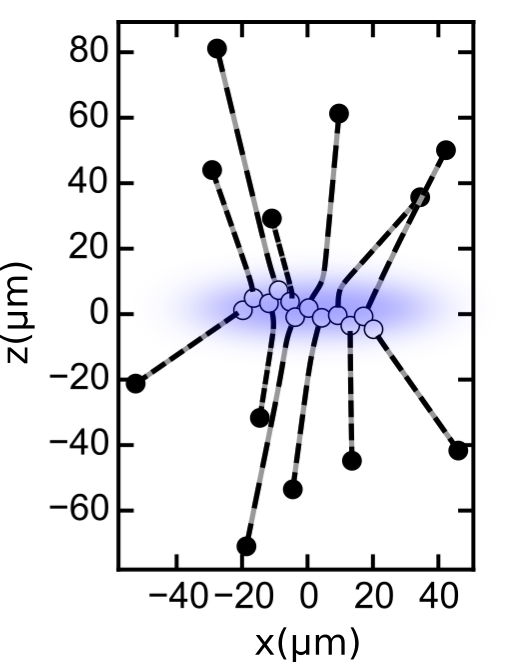}
\end{center}
\caption{Simulated van der Waals explosion in the $x-z$ plane of a 1D cluster of Rydberg atoms (white spheres represent the initial positions of the Rydberg atoms, and black spheres indicate the positions after $60\,\mathrm{\mu s}$). Each segment (black and white) of the trajectories corresponds to $5\,\mathrm{\mu s}$.}
\label{Fig4}
\end{figure}
\indent Fig. 3~(b) also shows the results of a numerical simulation of our experiment (see Fig. 4), which includes the van der Waals repulsion, the thermal motion of the atoms and the Coulomb repulsion after field ionization, which acts during the $\approx 10\,\mathrm{\mu s}$ time-of-flight towards the channeltron. The creation of the clusters is simulated by placing Rydberg atoms at a distance $r_{fac}$ with a probability weighted by the square of the Rabi frequency (since for our experimental parameters the excitation process is incoherent~\cite{Lesanovsky:13}) that varies perpendicularly to the $x$-direction. The constraint in the interparticle distance together with the spatially varying excitation probability gives rise to quasi-1D clusters that grow close to the $x$-axis with small (of the order of the width of the $420\,\mathrm{nm}$ beam) and randomly distributed radial separations from it. In our simulations the nuclear motion is treated classically with an adiabatic evolution of the electronic states and neglecting all nonadiabatic couplings~\cite{Ates:08}. For long expansion times of the clusters, Fig.3 (b) shows excellent agreement between the experiment and the numerical simulation without any adjustable parameters. Below $200\,\mathrm{\mu s}$ the simulation slightly overestimates the effect of the Coulomb explosion, which may be due to our incomplete knowledge of the exact trajectories of the individual ions. In the simulation a linear acceleration due to a perfectly homogeneous electric field along the $x$-axis was assumed, which is probably not an accurate representation given the geometry shown in Fig. 1 (these deviations will be investigated in detail in future work). The simulation also reveals that the deviation of the clusters from a perfect 1D configuration leads to a smaller expansion (by about a factor of 2) in the $x$-direction than in the radial directions. For hypothetical radial beam widths smaller than around $2\,\mathrm{\mu m}$, by contrast, the simulations show a larger expansion along the $x$-axis.\\
\indent In summary, we have measured the mechanical effect in real space of the vdW interaction between Rydberg atoms excited at close range. These measurements were made possible by interpreting the arrival time distributions of ions at the channeltron after field ionization of the Rydberg clusters.  Our observations indicate that the facilitated creation of Rydberg excitations at large values of the detuning, which leads to small separations between the excited atoms and hence to a large interaction energy, makes it possible to observe the van der Waals repulsion even in magneto-optical traps, without the need to go to lower temperatures such as those commonly reached in Bose-Einstein condensation. Whilst the present experiment, which projects the 3D Rydberg cloud onto a single time dimension, cannot observe details of the vdW interaction, that resolution can be achieved  in future experiments with 2D or 3D resolution based on large area imaging, precise multi-hit timing information and high detection rates, such as in the helium BEC experiment~\cite{Schellekens:05}.  Also, the Rydberg atoms accelerated as shown above could be used for controlled collision experiments. Finally, our results demonstrate the intrinsic mechanical instability of off-resonantly excited Rydberg gases, which has implications for proposed applications in quantum computation and quantum simulation \cite{Jaksch:00}. \\
\indent We acknowledge support by the European Union H2020 FET Proactive project RySQ (grant N. 640378) and the EU Marie Curie ITN COHERENCE.




\bibliographystyle{apsrev4-1}

\bibliography{BibliographyVdW_V11} 

\end{document}